 \documentstyle[emulateapj5]{aastex}

\newcommand{\mdot}{M$_{\odot}$}
\newcommand{\massu}{$\mathrm{M_U}$}
\newcommand{\edot}{$\dot{E}$}
\newcommand{\uopt}{$\mathrm{U_{opt}}$}
\newcommand{\hst}{{\it HST}}
\newcommand{\cha}{{\it Chandra}}
\newcommand{\teff}{$\mathrm{T_{eff}}$}

\slugcomment{Accepted for publication by {\it The Astrophysical Journal Letters}}

\shorttitle{Optical Detection of 47 Tuc MSP Companion}
\shortauthors{Edmonds et al.}

\begin{document}


\title{Optical Detection of a Variable Millisecond Pulsar Companion in
47 Tucanae \altaffilmark{1}}

\altaffiltext{1}{Based on observations with the NASA/ESA {\em Hubble Space
Telescope} obtained at STScI, which is operated by AURA, Inc. under NASA
contract NAS 5-26555.}


\author{Peter D. Edmonds\altaffilmark{2}}
\altaffiltext{2}{Harvard College Observatory, 60 Garden St, Cambridge, MA
02138; pedmonds@cfa.harvard.edu; cheinke@cfa.harvard.edu,
josh@cfa.harvard.edu}

\author{Ronald L. Gilliland\altaffilmark{3}}
\altaffiltext{3}{Space Telescope Science Institute, 3700 San Martin Drive,
Baltimore, MD 21218; gillil@stsci.edu}

\author{Craig O. Heinke and Jonathan E. Grindlay\altaffilmark{2}}

\author{Fernando Camilo\altaffilmark{4}}
\altaffiltext{4}{Columbia Astrophysics Laboratory, Columbia University, 550
West 120th Street, New York, NY 10027; fernando@astro.columbia.edu}



\begin{abstract}

Using results from radio and X-ray observations of millisecond pulsars in
47 Tucanae, and extensive \hst\ $U, V, I$ imaging of the globular cluster
core, we have derived a common astrometric solution good to $<$ 0\farcs1.
A close positional coincidence is found for 47 Tuc U, a 4.3\,ms pulsar in a
0.429 day orbit, detected in radio and X-rays, with an $m_V$ = 20.9 blue
star.  Analysis of extensive time series data for this optical candidate
shows a 0.004 magnitude semi-amplitude variation at the period and phase
expected from the radio ephemeris, and the optical variations are spatially
coincident with the candidate.  This provides secure optical detection of
the white dwarf companion to the millisecond pulsar, the first such
detection in a globular cluster, allowing for comparisons to recent models
for such companions with dependencies on mass and age.

\end{abstract}




\section{Introduction}

The huge improvement in spatial resolution of \cha\ over previous X-ray
missions is proving, like {\it Hubble Space Telescope} (\hst), to have
significant impact on globular cluster astrophysics. For example, using
\cha, 108 sources have been detected in the central 2$'\times2\farcm5$ of
the massive globular cluster 47 Tucanae (NGC104; Grindlay et
al. 2001\nocite{gri01}), compared to the detection of 9 central sources by
{\it ROSAT} \citep{ver98}. These \cha\ sources are a mixture of millisecond
pulsars (MSPs), cataclysmic variables (CVs), quiescent low-mass X-ray
binaries, and active binaries, many of which have also been detected
(Edmonds et al. 2001, in preparation) in our extensive 8.3 day \hst/WFPC2
observations of this cluster \citep{gil00}.  Fifteen of the 20 MSPs
detected in 47 Tuc by \citet{cam00} with the Parkes radio telescope have
timing positions accurate to $\lesssim$0\farcs01 (Freire et
al. 2001;\nocite{fre01} hereafter FCL01), and 10 of these are detected in
X-rays with the CIAO program {\tt wavdetect} (errors $\lesssim 0\farcs2$),
allowing for excellent astrometry between the radio and X-ray coordinate
frames. Combined with the large numbers of CVs and active binaries detected
with both \cha\ and \hst, we exploit here the rare opportunity to perform
astrometry at the $\sim$0.1$''$ level between X-ray, optical, and radio
observations, using only relative astrometry.

Of the 8 binary MSPs in 47 Tuc with timing solutions (FCL01\nocite{fre01}),
5 have companions with masses of $\sim $0.15--0.20\mdot\ \citep{cam00}; by
analogy with field MSPs these are expected to be helium white dwarfs (He
WDs). Using our precise astrometry we report here the unambiguous detection
of a He WD companion to one of these MSPs (PSR~J0024$-$7203U, hereafter 47
Tuc U), the first such detection in a globular cluster.  (He WDs have been
reported in the globular cluster NGC 6397 by Cool et
al. 1998\nocite{coo98}, Edmonds et al. 1999\nocite{edm99}, and Taylor et
al. 2001\nocite{tay01}, but none of these have yet been associated with an
MSP.)  The astrometry, photometry and time series for this identification
are described here, along with a discussion of their consequences for the
age of 47 Tuc U, the nature of He WDs and their cooling, and the heating
effects of an MSP in a close binary.

\section{Observations and Analysis}

\subsection{Astrometry}

We assumed the astrometric coordinate system of the MSPs (based on the JPL
DE200 planetary ephemeris; FCL01\nocite{fre01}) as our reference. First, we
transferred the \cha\ coordinate frame onto the radio frame using only the
relatively isolated MSPs with accurate \cha\ positions (six MSPs including
47 Tuc U). We calculated and subtracted the mean positional offsets between
the X-ray and radio coordinate systems and found the rms residuals were
$0\farcs136$ and $0\farcs037$ for RA and Dec respectively (after correcting
for small linear correlations between RA and RA error and Dec and Dec
error).  The total error of $0\farcs14$ compares favorably with the mean
predicted {\tt wavdetect} error, at the relevant count levels and in the
absence of background and crowding, of $0\farcs15$.

We then fixed the \hst\ coordinate system, with positions based on the
STSDAS program {\tt metric}, onto that of \cha\ using the X-ray detection
of 6 CVs and active binaries found on the WF2 chip of the
\citet{gil00} \hst/WFPC2 data (this chip contains 4 of the 6 MSPs,
including 47 Tuc U, used to perform the radio to \cha\ astrometry). The rms
residuals were found to be $0\farcs028$ and $0\farcs079$, for RA and Dec
respectively. Then, assuming the errors are random and dividing by
$\sqrt6$, the resulting 1$\sigma$ errors matching X-ray to optical are
$0\farcs011$ and $0\farcs032$ for RA and Dec, and for radio to optical are
$0\farcs057$ and $0\farcs049$.

\begin{figure}
\vspace*{-7cm}
\plotone{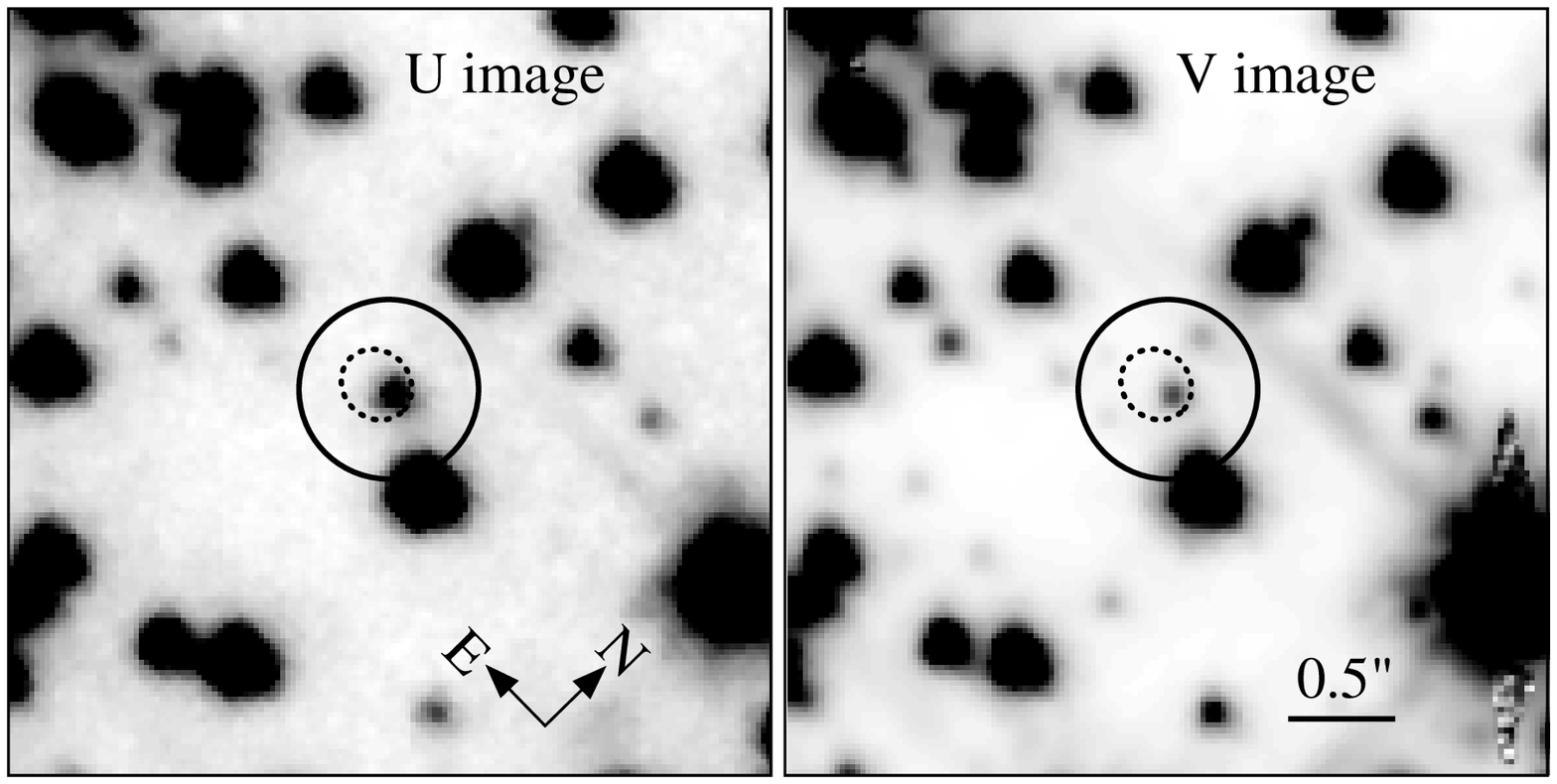}
\vspace*{-8cm}
\caption{Finding chart for 47 Tuc U showing the co-added (over 8.3 d),
oversampled images for the $U$- and $V$-band.  The solid line shows the
3$\sigma$ error circle (radius $=3\times0\farcs14$) for the X-ray source
W11, and the dotted line the 3$\sigma$ error ellipse for the radio pulsar
47 Tuc U, after accounting for the transformation between radio and optical
frames and assuming random errors. The optically detected binary companion
to 47 Tuc U (\uopt) is near the center of the W11 error circle.
\label{fig1}}
\end{figure}

Using this astrometry we searched directly for MSP companions in the \hst\
images and found an excellent candidate for a companion to 47 Tuc U
(hereafter \uopt) whose position is given in Table~1. Finding charts for
\uopt\ (Fig.~\ref{fig1}) show the $U$-band and $V$-band images, plus the
3$\sigma$ error circle for W11 (the \cha\ X-ray source corresponding to 47
Tuc U) and the 3$\sigma$ error ellipse for 47 Tuc U.  The offset between
W11 and \uopt\ is $0\farcs055$, and between 47 Tuc U and \uopt\ is
$0\farcs10$.

\subsection{Photometry}

Figure \ref{fig1} shows that \uopt\ is clearly blue, since other stars in
the $V$ image of comparable brightness to \uopt\ are barely visible in the
$U$ image.  This impression is confirmed by our $V$ vs $U-V$ and $V$ vs
$V-I$ color-magnitude diagrams (CMDs) from PSF-fitting photometry shown in
Figure \ref{fig2} (we have transformed from \hst\ filters to
Johnson-Cousins using Holtzman et al. 1995\nocite{hol95}; see also Table
1).  The star \uopt\ (labeled `U'; $M_{V}=7.47$) is blueward of the main
sequence (MS) in both CMDs, where the internal error for \uopt\ is
$\lesssim$ 0.03 mag for each filter, with absolute errors of $\sim0.05$ mag
in $V$ and $I$ and $\sim0.1$ mag in $U$, including transformation
uncertainties.  A likely CV is labeled (`1') as are 5 blue stars (circles)
confirmed by visual examination to be relatively bright in $U$ like
\uopt. One of these blue stars is probably a CO WD (see Zoccali et
al. 2001\nocite{zoc01}), two others are variable and likely CO WD-MS
binaries and the other two are possibly also binaries. Visual examination
showed that the other apparently blue stars, with $V < 22$, are caused by
crowding and photometric error, particularly in the crowded portions of the
image nearest to the cluster center. Many of the fainter stars with $V >
22$ and $0 \lesssim U-V \lesssim 1$ are bona-fide blue stars, likely to be
dominated by MS stars from the SMC (see Zoccali et al. 2001\nocite{zoc01}).

The very small number of stars that are blue and detectable in each of $U$,
$V$ and $I$ highlights the unusual nature of \uopt\ and its likely
association with 47 Tuc U. With 7 such blue stars the probability (averaged
over the whole WF2 image) of detecting one in a circular aperture of
$0\farcs15$ radius is only 9$\times 10^{-5}$.

\subsection{Time Series}

The \hst/WFPC2 data of \citet{gil00} contain 636 $V$-band and 653 $I$-band
160 s images evenly spread over 8.3 days. Typical rms noise values in
fractional intensity (Gilliland et al. 2000\nocite{gil00}) are $\sim0.003$
at $V=17$ and $\sim0.013$ at $V=20$. Since \uopt\ is relatively well
isolated, the quality of its $V$ and $I$ time series, based on its
extracted aperture of 0$\farcs34$, is excellent. Its rms noise values are
0.021 ($V$) and 0.034 ($I$), or $-0.1$ and 0.6$\sigma$ away from the mean
rms values for the 666 ($V$) or 551 ($I$) stars within 0.25 mag of \uopt,
after iteratively removing $>3\sigma$ outliers.

We then searched for evidence of low amplitude variations in \uopt.
Lomb-Scargle \citep{sca82} power spectra of the $V$ and $I$ time series for
47 Tuc U are shown in the upper panel of Figure \ref{fig3}. This shows a
peak in the $V$-band, at (2.681$\pm0.07)\times 10^{-5}$Hz, which is
coincident with the orbital period for 47 Tuc U ($2.697\times10^{-5}$Hz)
determined from radio observations. Although this peak is not statistically
significant in a blind search, only one other peak, when measured out to
0.45 mHz (the `average' Nyquist frequency; Scargle 1982\nocite{sca82}) has
a higher signal. No corresponding signal is seen in the $I$--band.
 
We then carried out a least-squares fit of a sinusoid to the time series,
by coordinating the radio ephemeris with the optical and correcting the
mid-exposure \hst\ times to the barycenter, using the IRAF task {\tt
rvcorrect}. We define phase$ = [T - (T_{asc} - P/4)]/P$ (where $T$ is the
barycentric time, $T_{asc}$ is the time of ascending node, when the pulsar
is moving with maximum velocity away from us, and $P$ is its orbital
period) so that when phase $= 0.0$ the heated hemisphere of \uopt\ is
facing towards us, giving maximum light (in a model where \uopt\ is heated
by the pulsar wind). The derived period, time of optical maximum (phase =
0) and semi--amplitude for the $V$-band time series are given in Table 1
($V$ phase plots are shown in the lower--left panel of
Fig.~\ref{fig3}). For the `unforced' solution the period is 0.7$\sigma$
away from the orbital radio period, and the phase is 1.9$\sigma$ away from
phase zero for the radio observations (as defined above). The amplitude is
significant at the 3.4$\sigma$ level (the significance of this measurement
is $>$ 3.4$\sigma$ when the $\sim$1/10 chance probability of the radio and
optical phases aligning is considered). By forcing the period and phase of
the sinusoidal variation to equal the radio values (the `forced' solution),
an amplitude of 0.0033$\pm$0.0012 is derived, significant at the
2.8$\sigma$ level; a forced fit in the $I$-band yielded 0.0013$\pm$0.0018.
Marginal evidence is found for variation in the much more limited (26
images of length 160 s -- 900 s) $U$-band data (see
Fig.~\ref{fig3}).

To test whether this small variation in $V$ is intrinsic to \uopt\ rather
than a larger variation from a nearby fainter star we examined the power
spectra of 3 fainter stars (with $V \gtrsim 22$) lying within
$\sim0\farcs5$ of \uopt. None showed significant variations. We also
examined a `phase difference' image derived by adding together images where
the phase of \uopt\ is within 0.07 of 1.0 and subtracting similar images
for phase values of 0.5$\pm$0.07. This procedure gave a stellar-shaped
source coincident, at the 0.5 pixel ($0\farcs05$) error level, with \uopt,
exactly as expected if \uopt\ is the only star varying with the radio
period and phase. We therefore conclude we have made a solid detection of
variability from \uopt\ at the expected period and phase, confirming it as
the 47 Tuc U companion.

\section{Discussion}

Neutron stars (NSs) in binary radio pulsar systems have average masses of
$\mathrm{M_{NS}}=1.35\pm0.05$\,\mdot\ \citep{tho99}.  For 47~Tuc~U,
considering a NS mass 1.30\,\mdot\ and an edge-on orbit (inclination $i =
90$\degr), we derive from the mass function a minimum mass for \uopt\ of
\massu\ = 0.120\,\mdot.  This 4.3\,ms pulsar has an approximate
characteristic age of $\tau \sim 2$ Gyr and spin-down luminosity of
\edot\ $\sim 2\times10^{34}$ erg s$^{-1}$ (Grindlay et al. 2001, in
preparation).  Such recycled pulsars may have accreted significant amounts
of mass, and for every additional 0.05\,\mdot\ for the NS, \massu's minimum
value increases by 0.003\,\mdot.

We now compare the $U, V, I$ colors of \uopt\ with expectations from H and
He WD models to test whether it is a plausible He WD and if so, to estimate
\teff, log g, and the mass and age of this star. We focus on comparisons
with the models of Serenelli et al. (2001; hereafter SAR01\nocite{ser01}),
who conveniently provide broadband color information.  They find that for
masses ranging from 0.18 \mdot\ to 0.41 \mdot, diffusion-induced
hydrogen-shell flashes take place which deplete the H envelope and cause
evolution to proceed relatively quickly, while the He WDs with mass
$<0.18$\mdot\ evolve much more slowly because they retain relatively thick
hydrogen envelopes. \citet{alt01} and SAR01\nocite{ser01} show that this
dichotomy between low and high mass He WDs removes the discrepancies
between MSP characteristic ages and WD cooling ages for several field
systems including PSRs J0034$-$0534, J1012+5307, J1713+0747 and B1855+09.

\begin{figure}
\vspace*{-7cm}
\plotone{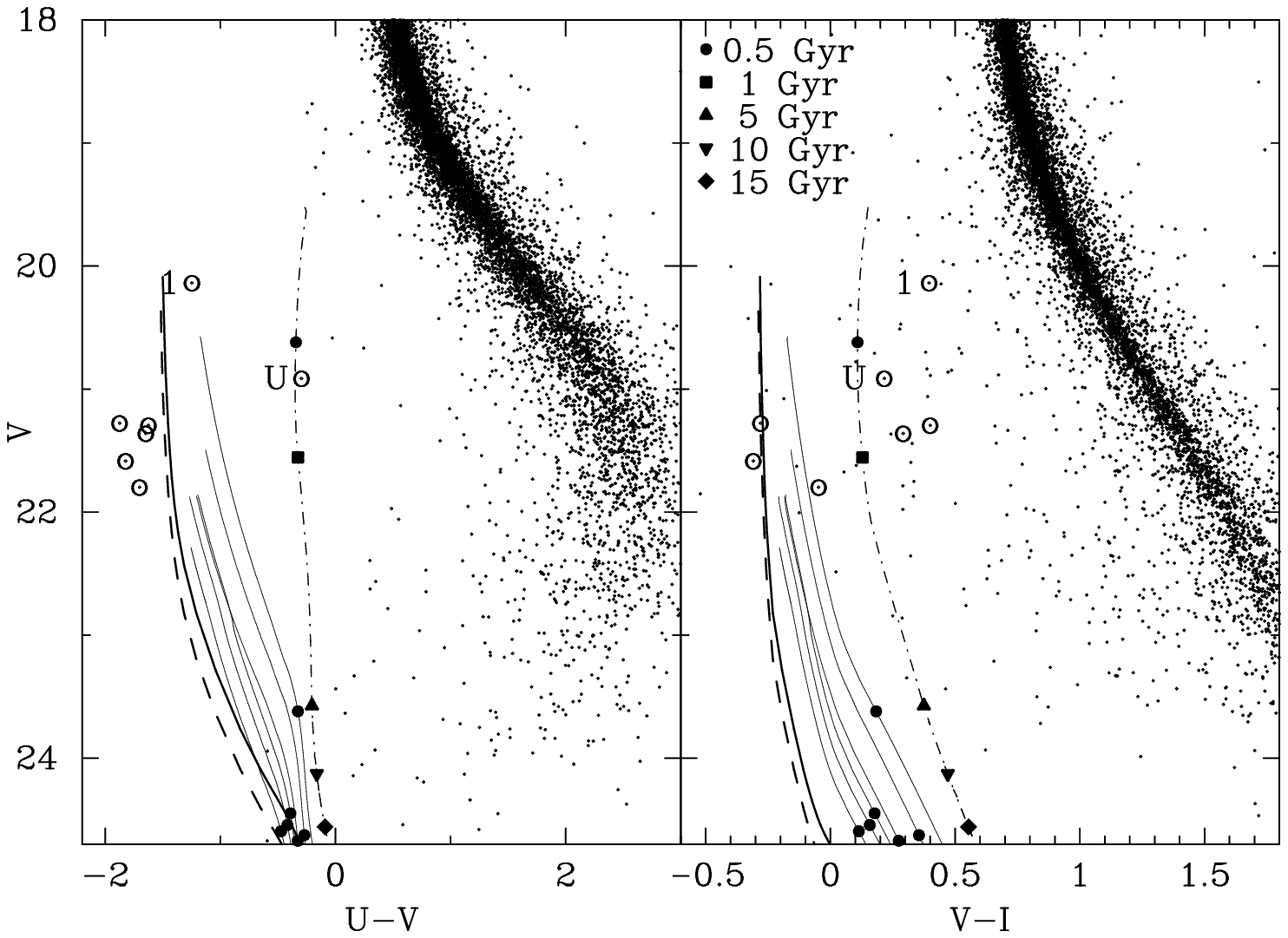}
\vspace*{-7cm}
\caption{$V$ vs $U-V$ and $V$ vs $V-I$ CMDs for the WF2 chip. The 47~Tuc~U
MSP companion \uopt\ is denoted by `U', a CV by `1' and other blue stars
are circled.  The position of \uopt\ is slightly redder and brighter than
those of the He WDs in NGC 6397 reported in \citet{coo98}.  The thicker
lines show \citet{ber95} models for 0.5 \mdot\ (dashed line) and 0.6 \mdot\
(solid line) CO WDs. The thinner lines are SAR01 models, for He WD masses
of 0.406, 0.360, 0.327, 0.292, 0.242, 0.196 \mdot, with mass decreasing
towards the red, and 0.169 \mdot\ (dot-dashed line). Cooling ages are as
shown.  These models have been plotted assuming that $(m-M)_0=13.27$, the
mean of the 9 distance modulus values reported in \citet{zoc01},
$E(B-V)=0.055$ \citep{zoc01}, $A_U/A_V=1.51$ and $A_I/A_V=0.60$
\citep{hol95}. Using $(m-M)_0=13.27$ the 47 Tuc distance is 4.5
kpc. \label{fig2}}
\end{figure}

\begin{figure}
\vspace*{-7cm}
\plotone{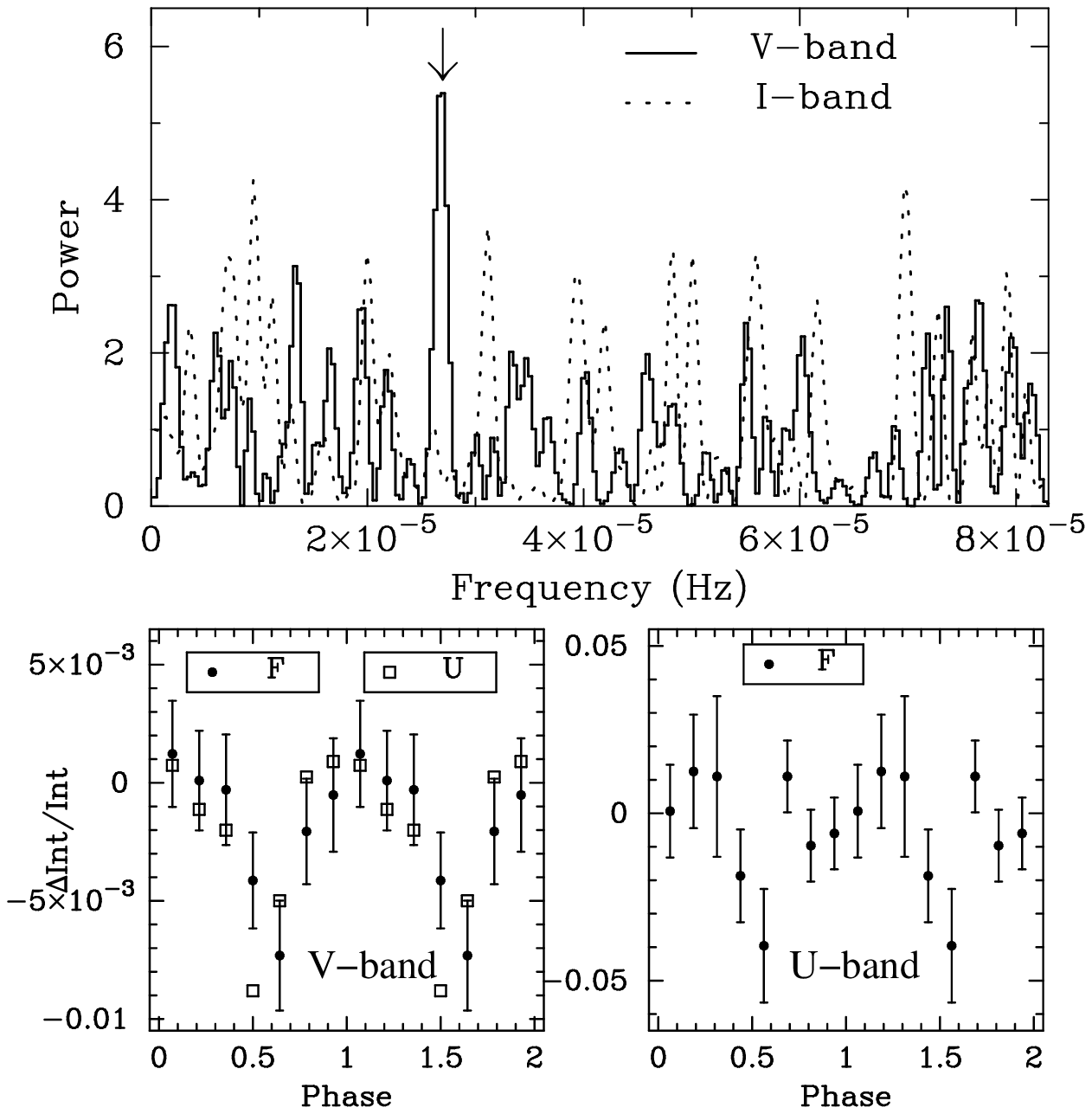}
\vspace*{-7.5cm}
\caption{Power spectra and phase plots for \uopt.  The upper panel shows
power spectra ($V$- and $I$-band, as labeled) for \uopt, with power shown
in Lomb-Scargle units \citep{sca82}. The binary frequency of 47 Tuc U
measured from radio observations is indicated with an arrow.  The lower
panel shows phase plots for the $V$- and $U$-band data for \uopt, where
fractional intensities are plotted and the errors are from the time series
rms values scaled by the square root of the number of points in each
bin. `F' denotes the forced fit and `U' the unforced fit (see \S
2.3 for description).  The point near phase 0.5 in the $U$-band data is
formally only significant at the 2.3$\sigma$ level. No evidence is found
for variation in the $I$-band, and so the phase plot is not shown here.
\label{fig3}}
\end{figure}

Figure \ref{fig2} shows the He WD cooling curves of SAR01\nocite{ser01} for
masses of 0.169 -- 0.406 \mdot.  Since \uopt\ lies well to the red of the
WD cooling sequence of \citet{ber95} in both CMDs, it is unlikely to be a
$\sim$0.55 \mdot\ CO WD. The lowest mass models of SAR01\nocite{ser01}
offer much better agreement, and since \uopt\ falls slightly to the red of
the lowest mass sequence we infer that \massu\ $\lesssim 0.17$\mdot\
(implying $i \gtrsim45$\degr). This would mean that 47 Tuc U is similar to
the field system PSR J1012+5307, with period = 0.605\,d and mass
$\sim0.16$\,\mdot\ \citep{cal98}.

The 0.17\mdot\ SAR01\nocite{ser01} model with the same absolute magnitude
as \uopt\ has \teff=11000 K, log g = 5.6, luminosity = 0.14 $L_{\odot}$, a
radius\footnote{The estimated radius is fairly insensitive to the details
of the \citet{ser01} WD models, since \teff\ can be estimated accurately
just from the $V-I$ color \citep{han98}, and the luminosity using only the
$V$ magnitude and \teff.} of $\sim7.3 \times10^{9}$cm and an age of
$\sim0.6$ Gyr. This cooling age is somewhat less than the characteristic
age estimated by Grindlay et al. (2001, in preparation) and is at the low
end of the conservative limits set by FCL01\nocite{fre01} of $0.4 < \tau <
4.2$ Gyr. Only the lowest mass model of SAR01\nocite{ser01} remains
relatively bright at these significant cooling ages, while all of the
higher mass WDs reach much fainter levels ($M_V \sim12$) at a cooling age
of $\sim1$ Gyr (see Fig.~\ref{fig2}). Our data therefore provides tentative
support for the evolutionary calculations of SAR01\nocite{ser01}, in
particular for the dichotomy in ages between different mass He WDs.  For
comparison, the He WD models of \citet{han98} (for all masses) reach
$M_I$=11.5 or fainter after 1 Gyr, for a relatively thick hydrogen envelope
of $3\times10^{-4}$\,\mdot.

Accretion should have ceased in the 47 Tuc U system, to explain the faint
X-ray luminosity ($2\times10^{30}$erg s$^{-1}$; Grindlay et al. 2001, in
preparation) and the constant, uneclipsed radio emission
(FCL01\nocite{fre01}). Using the Roche-lobe formula from \citet{pac71}
($r/a=\mathrm{0.462[(M_{U}/(M_{NS}+M_U)]^{1/3}}$, where $r$ is the
Roche-lobe radius, and $a$ is the binary separation), the WD radius given
above (for \massu\ = 0.17 \mdot; $i = 45$\degr) and the binary separation
from Kepler's Third Law, \uopt\ should under--fill its Roche lobe by a
factor of $\sim$6. Therefore, as expected, no accretion will be occurring.

%This is perhaps not surprising given the uncertainties
%involved (e.g. 47 Tuc has a lower metallicity than the solar metallicity
%assumed in the SAR01\nocite{ser01} models). 

Assuming that the MSP radiates its wind isotropically, and that $i =
45$\degr, then the \edot\ estimate implies that the pulsar energy
intercepted by \uopt\ should be $\sim7 \times10^{30}$erg s$^{-1}$ or
0.013$\times$ the luminosity of \uopt\ (assuming that the spin-down energy
is re-radiated as a black-body, we derive a temperature of $\sim5200$K,
peaked at $\sim5600$\AA, for the re-radiated emission).  Therefore,
assuming that only one side of \uopt\ is heated we should see a relative
variability amplitude of $\sim0.005$, allowing a factor of 1.4 decrease in
possible amplitude because of the inclination. This is only slightly larger
than the amplitude of the detected $V$-band variation, implying a high
efficiency factor for re-radiation of the pulsar's wind as optical
emission.  However, because the variability expected from the effects of
heating is small, we infer that the possible (2.3$\sigma$) much larger
amplitude variation in the $U$--band is either spurious or evidence for a
separate source of emission.

\section{Summary and Followup Observations}

We have identified the binary pulsar 47 Tuc U with an optically variable
blue star with period and phase of maximum light in excellent agreement
with that expected from the precise radio ephemeris (FCL01\nocite{fre01}).
The companion is likely to be a He white dwarf with \massu\ $\lesssim$0.17
\mdot, based on the published models of SAR01\nocite{ser01}.

Although the optical time series photometry reported here is unlikely to be
improved upon soon, time resolved spectroscopy with \hst/STIS of \uopt\
appears promising. The co-added Doppler corrected spectrum could be used to
spectroscopically determine \teff\ and log g, and hence \massu\ (see
e.g. Edmonds et al. 1999\nocite{edm99}).  Combining \massu\ with the radial
velocity amplitude of \uopt\ ($\mathrm{K_U}$) and the measured
$\mathrm{K_{NS}}$ from the radio measurements would give a direct measurement
of $\mathrm{M_{NS}}$.

%% In this section, we use  the \subsection command to set off
%% a subsection.  \footnote is used to insert a footnote to the text.

%% cross-references to sections, equations, tables, and figure

%% This section also includes several of the displayed math environments
%% mentioned in the Author Guide.

\acknowledgments

We are grateful for discussions and modeling help from Leandro Althaus.  We
thank Ata Sarajedini, Raja Guhathakurta, and Justin Howell for contributing
to the photometric analysis, Hans Ritter and Andrew King for helpful
comments on the manuscript and Fred Rasio, Vicky Kalogera, Kailash Sahu for
discussions.  This work was supported in part by STScI grant GO-8267.01-97A
(PDE and RLG) and by NASA grant NAG 5-9095 (FC).

\clearpage 

%---------------------------------------------------------------------------

%\documentclass[preprint]{aastex}
%\begin{document}

\begin{deluxetable}{cccccccc}
\tablecolumns{8}
\tablewidth{0pc}
\tablecaption{Positional, photometric and time series information for 47
Tuc U and its optical companion \uopt}
\tablehead{
 \colhead{RA\tablenotemark{a}} & \colhead{Dec\tablenotemark{a}} & 
 \colhead{$U$} & \colhead{$V$} & \colhead{$I$} & \colhead{period$(V)$} &
 \colhead{$T$(phase=0)\tablenotemark{b}} & \colhead{amp$(V)$} \\
\colhead{(J2000)} & \colhead{(J2000)} & \colhead{} & \colhead{} & \colhead{} & 
\colhead{(days)} & \colhead{(MJD)} & \colhead{(mag)} 
}
\startdata

00 24 09.8325(5) & $-$72 03 59.667(3) & 20.6(1) & 20.91(5) & 20.70(5) &
0.432(4) & 51367.32(2) & 0.004(1) 
             
\tablenotetext{a}{Coordinates from Freire et al. (2001)}
\tablenotetext{b}{Time of optical maximum (see text)}

%\tablecomments{bla bla bla}

\enddata
\end{deluxetable}

%\end{document}

%---------------------------------------------------------------------------

\clearpage


\clearpage

\clearpage

%% If you are not including electonic art with your submission, you may
%% mark up your captions using the \figcaption command. See the 
%% No more than seven \figcaption commands are allowed per page, 
%% so if you have more than seven captions, insert a \clearpage 

%% Tables should be submitted one per page, so put a \clearpage before

%% Two options are available to the author for producing tables:  the
%% deluxetable environment provided by the AASTeX package or the LaTeX
%% table environment.  Use of deluxetable is preferred.
%%

%% Three table samples follow, two marked up in the deluxetable environment,
%% one marked up as a LaTeX table.

%% In this first example, note that the \tabletypesize{}
%% command has been used to reduce the font size of the table.
%% Note also that the \label command needs to be placed 

\end{document}